\documentclass[twocolumn,twoside]{IEEEtran}

\ifCLASSINFOpdf

\else

\fi

\ifCLASSOPTIONcompsoc
\usepackage[caption=false, font=normalsize, labelfont=sf, textfont=sf]{subfig}
\else
\usepackage[caption=false, font=normalsize]{subfig}
\fi
\usepackage{lipsum}%
\usepackage[dvipsnames]{xcolor}

\usepackage{balance}
\usepackage{multicol}   
\usepackage{cite}
\usepackage{gensymb}
\usepackage{multirow}
\usepackage{graphics}
\usepackage{epsfig}
\usepackage{graphicx}
\usepackage{epstopdf}
\usepackage{textcomp}
\usepackage{amsmath}
\usepackage{mathtools}
\usepackage{filecontents}
\usepackage{lipsum,color}
\usepackage{amssymb}
\usepackage{float}
\usepackage{comment}

\usepackage{amsthm}

\usepackage{braket}
\usepackage{times} 
\usepackage{amsthm}  

\usepackage{amsfonts}

\theoremstyle{break}

\begin{document}
\title{Hierarchical Deep Learning for Joint Turbulence and PE  Estimation in Multi-Aperture FSO Systems}

\author{Mohammad~Taghi~Dabiri,~Meysam~Ghanbari,
	~Rula~Ammuri, \\
	~Mazen~Hasna,~{\it Senior Member,~IEEE}, 
	~and~Khalid~Qaraqe,~{\it Senior Member,~IEEE}

	\thanks{M.T. Dabiri, and M. Ghanbari are with the Qatar Center for Quantum Computing, College of Science and Engineering, Hamad Bin Khalifa University, Doha, Qatar. email: (mdabiri@hbku.edu.qa; megh89467@hbku.edu.qa; smalkuwari@hbku.edu.qa).}
	
	\thanks{Rula Ammuri is with Professionals for Smart Technology (PST), Amman, Jordan (email: rammuri@pst.jo).}
	
	\thanks{Mazen Hasna is with the Department of Electrical Engineering, Qatar University, Doha, Qatar (e-mail: hasna@qu.edu.qa).}
	
	\thanks{Khalid A. Qaraqe is a professor with the College of Science and Engineering, Hamad Bin Khalifa University, Doha, Qatar, and  an adjunct  professor with the Department of Electrical Engineering, Texas A\&M University at Qatar, Doha, Qatar (e-mail: kqaraqe@hbku.edu.qa)}
	
	\thanks{This publication was made possible by NPRP14C-0909-210008 from the Qatar Research, Development and Innovation (QRDI) Fund (a member of  Qatar Foundation). The statements made herein are solely the responsibility of the author[s].}
}

\maketitle
\begin{abstract}
Accurate characterization of free-space optical (FSO) channels requires joint estimation of transmitter pointing errors, receiver angle-of-arrival (AoA) fluctuations, and turbulence-induced fading. However, existing literature addresses these impairments in isolation, since their multiplicative coupling in the received signal severely limits conventional estimators and prevents simultaneous recovery. In this paper, we introduce a novel multi-aperture FSO receiver architecture that leverages spatial diversity across a lens array to decouple these intertwined effects. Building on this hardware design, we propose a hierarchical deep learning framework that sequentially estimates AoA, transmitter pointing error, and turbulence coefficients. This decomposition significantly reduces learning complexity and enables robust inference even under strong atmospheric fading. Simulation results demonstrate that the proposed method achieves near-MAP accuracy with orders-of-magnitude lower computational cost, and substantially outperforms end-to-end learning baselines in terms of estimation accuracy and generalization. To the best of our knowledge, this is the first work to demonstrate practical joint estimation of these three key parameters, paving the way for reliable, turbulence-resilient multi-aperture FSO systems.
\end{abstract}

%

%
\IEEEpeerreviewmaketitle


\section{Introduction}
The growing demand for high-capacity, secure, and flexible connectivity in next-generation wireless and space–air–ground networks has accelerated interest in free-space optical (FSO) communication. FSO is widely considered for applications such as backhaul extension, unmanned aerial vehicle (UAV) to ground links, inter-satellite communication, and quantum key distribution (QKD), making it a strong candidate for future communication infrastructures \cite{kumar2025skr,xu2025multi}. 
Despite its potential, FSO is highly sensitive to atmospheric turbulence, transmitter pointing error (PE), and receiver angle-of-arrival (AoA) fluctuations. These effects are particularly severe in mobile and aerial scenarios where maintaining beam stability is challenging. Apart from inter-satellite links that are essentially turbulence-free, most terrestrial and aerial FSO systems employ dedicated tracking subsystems that operate independently from the communication link. A more scalable solution is to enable \textit{joint communication and tracking} within the same receiver architecture so that channel impairments are estimated and mitigated concurrently without extra hardware \cite{10681506,10553228}. To address such impairments, various receiver-side strategies have been investigated: aperture averaging \cite{paul2022pulse}, multi-aperture diversity receivers \cite{girdher2024ris}, and adaptive optics (AO) \cite{ata2023haps} for turbulence, while beam tracking systems are commonly used to counter pointing jitter \cite{solanki2025computationally}. The effectiveness of all these techniques, however, critically depends on accurate estimation of pointing offsets and AoA, since residual misalignment or uncorrected distortions directly reduce link reliability.

A maximum-likelihood framework with expectation–maximization (EM) has been shown to outperform Newton–Raphson (NR) and generalized method of moments (GMM) estimators, achieving accuracy close to the Cramér–Rao bound (CRB) under both simulated and experimental conditions \cite{chen2021parameter}. Learning-based estimators, such as adaptive incremental networks, have also been applied under log-normal and Gamma–Gamma turbulence, highlighting the increasing difficulty of accurate tracking as turbulence strengthens \cite{abdavinejad2020fso}. More recently, refined moment-based techniques that explicitly incorporate noise have delivered robust performance at low signal-to-noise ratio (SNR) and have outperformed conventional moment methods in strong turbulence regimes \cite{kim2025moment}.

The system-level impact of pointing uncertainty has been quantified, showing that imperfect channel state information can significantly degrade outage probability, bit-error rate (BER), and capacity \cite{han2022joint}. For real-time inference, extended Kalman filtering (EKF) has been used for onboard pointing-error detection in satellite systems, reducing reliance on ground tracking and improving orbit and attitude determination \cite{dhanalakshmi2022onboard}. In addition, saddlepoint approximation (SAP) methods for generalized pointing-error models provide accurate parameter estimation in both noiseless and noisy conditions and are suitable for long-distance FSO links \cite{miao2021parameter}.

Angle-of-arrival (AoA) inference has received increasing attention in mobile FSO scenarios. Beam-shape methods, including ellipse fitting and Gaussian least-squares, have been validated against the CRB and shown to be reliable under noise and discretization effects \cite{dreier2025beam}. Complementary techniques that exploit both spot location and received energy for narrow Gaussian beams demonstrate that incorporating energy information markedly improves accuracy for small AoA values and narrow beamwidths, even in the presence of random pointing errors \cite{tsai2023angle}.

Taken together, these studies demonstrate substantial progress when each impairment is considered in isolation. In practice, however, atmospheric turbulence, pointing errors, and AoA fluctuations occur simultaneously, and their effects mix multiplicatively in the received signal. This coupling makes it difficult to isolate and accurately estimate the underlying parameters with conventional analytical tools. Because most prior work treats these factors separately, the combined dynamics of mobile and aerial FSO links are not fully captured. This gap motivates unified \emph{joint estimation} frameworks that address the composite effect of turbulence, pointing error, and AoA within a single model, thereby providing more reliable inputs to downstream mitigation strategies and improving link robustness in practical deployments.

We propose a multi-aperture FSO receiver that replaces a single large optic with a lens array feeding quad photodiodes, and we develop a tractable end-to-end model that includes per-lens pointing gain and closed-form quad-cell AoA gains via a separable Gaussian PSF and normal CDFs together with Gamma–Gamma turbulence. Building on this, we design a hierarchical deep-learning estimator: Stage 1 infers AoA from normalized quad ratios to suppress common multiplicative scaling; Stage 2 estimates the transmitter pointing error (PE) from AoA-compensated per-lens aggregates; Stage 3 recovers receiver jitter by subtraction and obtains per-aperture turbulence by multiplicative pointing removal. Compared with exhaustive maximum a posteriori (MAP) search, the method attains near-MAP accuracy with orders-of-magnitude lower complexity and low latency; relative to end-to-end learning, it improves AoA accuracy and data efficiency while remaining competitive for PE and enabling reliable turbulence recovery once PE is estimated. Extensive simulations across array sizes and turbulence strengths quantify NMSE trends and yield practical sizing guidelines, and a physics-grounded data-generation pipeline enables reproducible training and evaluation. To our knowledge, this is the first practical joint estimation of AoA, PE, and per-aperture turbulence in a multi-aperture FSO receiver using a hierarchical deep-learning framework.

\begin{figure}
	\centering
	\subfloat[] {\includegraphics[width=3.3 in]{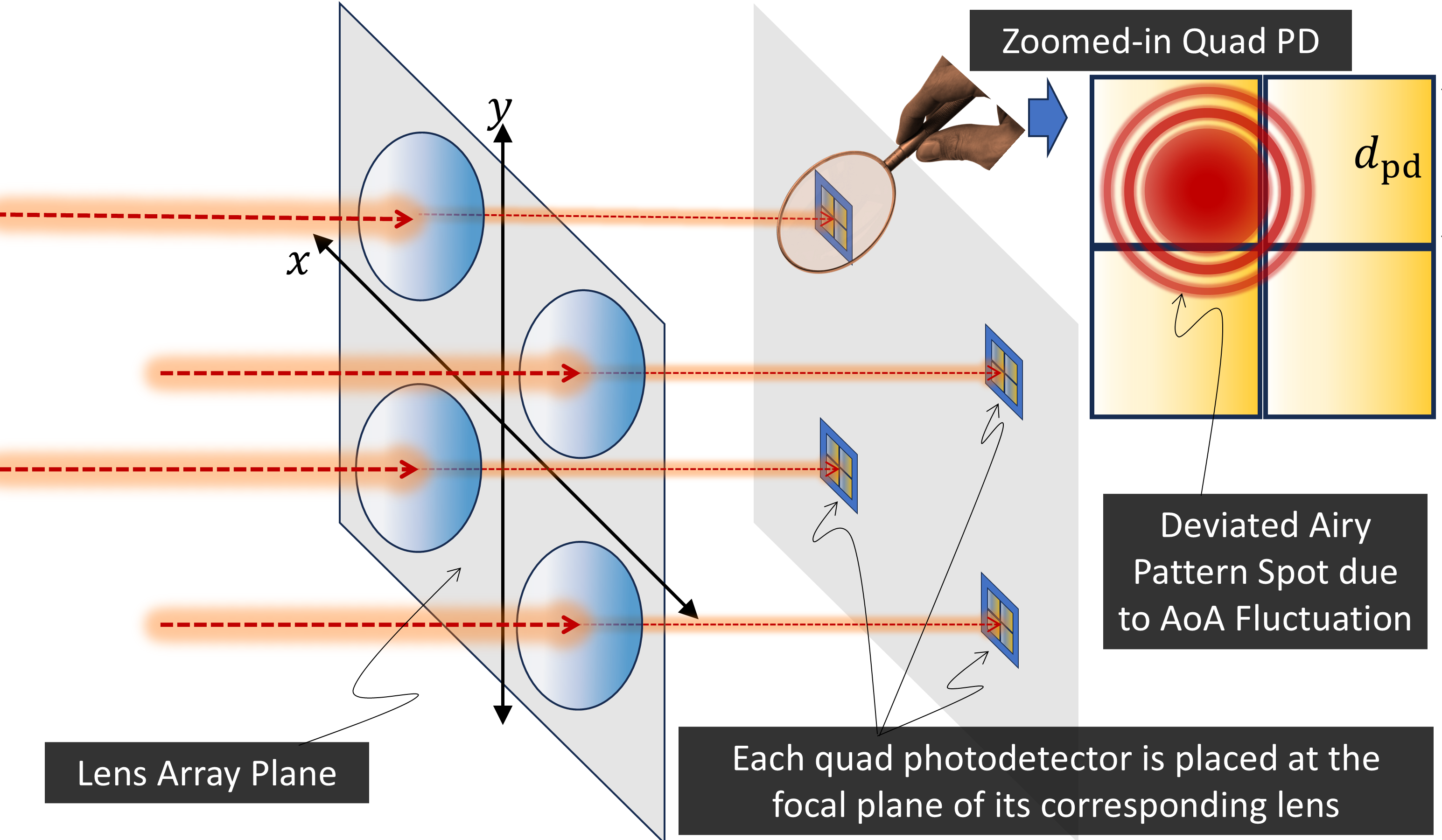}
		\label{cz1}
	}
	\hfill
	\subfloat[] {\includegraphics[width=2.2 in]{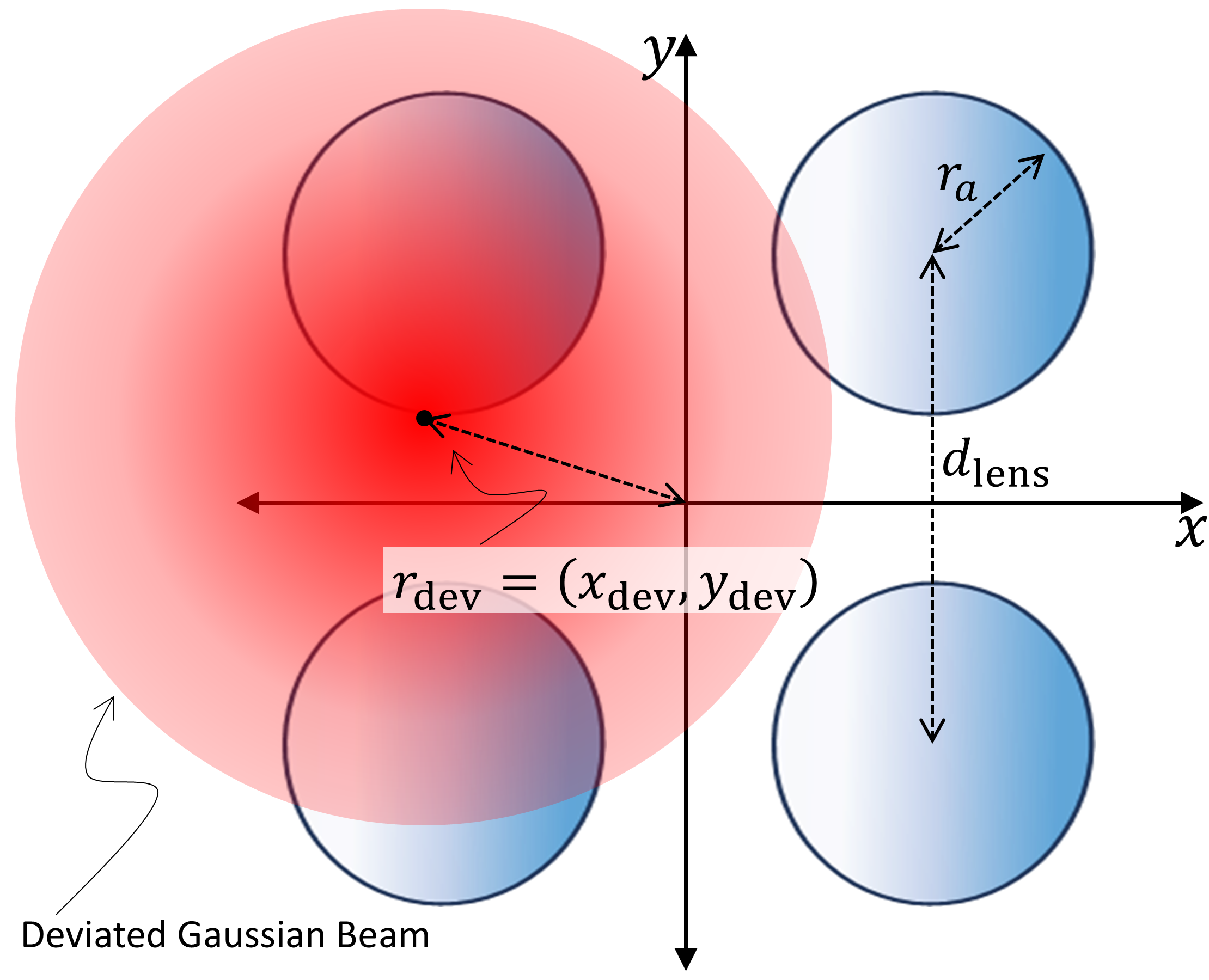}
		\label{cz2}
	}
	\caption{(a) Schematic of the proposed multi-aperture FSO receiver, where an array of several small lenses is employed instead of a single large lens. Each lens focuses the incoming beam onto a quad PD located at its focal plane; the zoomed-in panel shows the Airy-spot displacement across the quad PD due to AoA fluctuations.
	(b) Geometric model of the deviated Gaussian beam on the lens plane, where $r_{\mathrm{dev}}=(x_{\mathrm{dev}},y_{\mathrm{dev}})$ is the beam-deviation vector, $r_a$ is the lens aperture radius, and $d_{\mathrm{lens}}$ denotes the inter-lens spacing (pitch).}
	
	\label{cz}
\end{figure}

\section{System Model}
Fig.~\ref{cz}(a) illustrates the overall architecture of the proposed multi-aperture FSO receiver. 
Instead of employing a single large lens, an array of multiple smaller lenses is used. 
This design leads to a shorter focal length, a reduced physical footprint, and, importantly, 
a higher robustness against pointing errors and atmospheric turbulence. Each small lens focuses the incident beam 
onto its dedicated quad photodetector (PD), enabling spatially resolved measurements that 
are later exploited for angle-of-arrival and turbulence estimation.

As illustrated in Fig.~\ref{cz}(b), the lens array is located on the $x$–$y$ plane, 
while the $z$-axis represents the optical propagation direction from transmitter to receiver. 
The link distance is denoted by $Z_L$. The transmitter points the optical beam 
towards the receiver with a divergence angle $\theta_{\mathrm{div}}$. 
Due to tracking imperfections, the beam axis may deviate from the ideal propagation axis, 
resulting in a pointing error 
\begin{align}
	\boldsymbol{\theta}_e = (\theta_{ex},\theta_{ey}),
\end{align}
where $\theta_{ex},\theta_{ey}\sim\mathcal{N}(0,\sigma_\theta^2)$ are modeled as zero-mean Gaussian random variables. 
Consequently, the center of the Gaussian beam at the receiver plane is shifted by
\begin{align}
	\mathbf{r}_{\mathrm{dev}}(t) 
	&= \begin{bmatrix}x_{\mathrm{dev}}(t) \\ y_{\mathrm{dev}}(t)\end{bmatrix} 
	= Z_L \cdot \boldsymbol{\theta}_e(t),
	\label{eq:dev}
\end{align}
where $\boldsymbol{\theta}_e=(\theta_{ex},\theta_{ey})$ with 
$\theta_{ex},\theta_{ey}\sim\mathcal{N}(0,\sigma_\theta^2)$, 
and thus $\mathbf{r}_{\mathrm{dev}}(t)\sim\mathcal{N}(\mathbf{0},(Z_L\sigma_\theta)^2\mathbf{I}_2)$.

The received optical intensity distribution on the $x$–$y$ plane is modeled as a 
circularly symmetric Gaussian beam
\begin{align}
	I(x,y) &= I_0 \exp\!\left(-\frac{(x-x_{\mathrm{dev}})^2+(y-y_{\mathrm{dev}})^2}{w_z^2}\right),
	\label{eq:gauss}
\end{align}
where $I_0$ is the peak irradiance and $w_z$ is the beam waist at distance $Z_L$.
The lens array consists of $N_{\text{lens}}$ small apertures. 
For illustration, Fig.~\ref{cz}(a) shows the case $N_{\text{lens}}=4$. 
The center of lens $i$ is located at $\mathbf{p}_i=(x_i,y_i)$ 
with aperture radius $r_a$, where $i\in\{1,...,N_{\text{lens}}\}$. 
The received optical power collected by lens $i$ is obtained as
\begin{align}
	P_i &= \iint_{\|\mathbf{r}-\mathbf{p}_i\|\leq r_a} I(\mathbf{r}) \, d\mathbf{r} ,
	\label{eq:pointing}
\end{align}
which corresponds to the pointing gain factor of lens $i$.
For practical design, the beam waist is typically much larger than the lens radius 
($w_z \gg r_a$), so the intensity across each lens can be approximated as constant.
Under the condition $w_z \gg r_a$, the Gaussian spot intensity is approximately 
constant over the aperture of each lens. Therefore, the integral in 
\eqref{eq:pointing} reduces to
\begin{align}
	P_i &\approx I(\mathbf{c}_i)\,\pi r_a^2 ,
	\label{eq:pointing_simple}
\end{align}
where $I(\mathbf{c}_i)$ is the beam intensity evaluated at the center of lens $i$.

The pointing gain coefficient of lens $i$ is obtained by normalizing the collected 
power $P_i$ with respect to the maximum (on-axis) collected power. Using 
\eqref{eq:dev}–\eqref{eq:pointing_simple}, it can be expressed as
\begin{align}
	&h_{p,i}(\theta_{ex},\theta_{ey}) 
	= \frac{r_a^2}{w_z^2} 
	\times e^{-\frac{(x_i - Z_L \theta_{ex})^2 
		+ (y_i - Z_L \theta_{ey})^2}{w_z^2}}.
	\label{eq:hpi_full}
\end{align}

\subsection{AoA Modeling}
In addition to the transmitter-side tracking error $\boldsymbol{\theta}_e=(\theta_{ex},\theta_{ey})$, 
mobile platforms (e.g., UAVs) introduce small random tilts of the lens-array plane due to vibration. 
We denote this receiver-induced angular jitter by 
$\boldsymbol{\theta}_r=(\theta_{rx},\theta_{ry})$, modeled as a zero-mean Gaussian process \cite{dabiri2025novel}
\begin{align}
	\boldsymbol{\theta}_r &\sim \mathcal{N}\!\big(\mathbf{0},\,\sigma_r^2 \mathbf{I}_2\big),
	\label{eq:theta_r_dist}
\end{align}
independent of $\boldsymbol{\theta}_e$. 
The instantaneous \emph{angle-of-arrival (AoA)} is then defined as the sum of two independent Gaussian vectors:
\begin{align}
	\boldsymbol{\theta}_{\mathrm{AoA}}=(\theta_{\mathrm{AoA,x}},\theta_{\mathrm{AoA,y}})
	&= \boldsymbol{\theta}_e + \boldsymbol{\theta}_r,
	\label{eq:aoa_def}
\end{align}
which yields the effective AoA distribution
\begin{align}
	\boldsymbol{\theta}_{\mathrm{AoA}} &\sim \mathcal{N}\!\big(\mathbf{0},\,\sigma_\theta^2\mathbf{I}_2+\sigma_r^2\mathbf{I}_2\big)
	= \mathcal{N}\!\big(\mathbf{0},\,\sigma_{\theta,\mathrm{eff}}^2\mathbf{I}_2\big),
	\label{eq:aoa_gauss}
\end{align}
where $\sigma_{\theta,\mathrm{eff}}^2 \triangleq \sigma_\theta^2+\sigma_r^2$.

After each lens, the incoming angle-of-arrival (AoA) produces a lateral shift of the focal-spot on the quad PD. 
Under the small-angle approximation, the spot displacement on the focal plane of lens $i$ is
\begin{align}
	\mathbf{s}_i 
	&= \begin{bmatrix} s_{i,x} \\ s_{i,y} \end{bmatrix}
	= f_c \, \tan(\boldsymbol{\theta}_{\mathrm{AoA}})
	\simeq f_c \,\boldsymbol{\theta}_{\mathrm{AoA}}
	= f_c \begin{bmatrix} \theta_{\mathrm{AoA},x} \\ \theta_{\mathrm{AoA},y} \end{bmatrix},
	\label{eq:spot_shift}
\end{align}
where $f_c$ is the focal length of the small lens (common across lenses in the array).

The point-spread function (PSF) on the focal plane is approximated by a
circularly symmetric Gaussian distribution with effective spot width
$w_{\text{spot}}$. Let $\mathbf{r}=[x_f,y_f]^\top$ denote a focal-plane
coordinate referenced to the optical axis of lens $i$. The irradiance at
the focal plane of lens $i$ is expressed as
\begin{align}
	I^{(i)}_f(\mathbf{r})
	&= \frac{1}{\pi w_{\text{spot}}^2}
	\exp\!\left(
	-\frac{\|\mathbf{r}-\mathbf{s}_i\|^2}{w_{\text{spot}}^2}
	\right),
	\label{eq:gaussian_psf}
\end{align}
where $\mathbf{s}_i$ is the AoA-induced spot displacement from
\eqref{eq:spot_shift}. The parameter $w_{\text{spot}}$ represents the
effective beam waist at the focal plane, chosen such that the Gaussian
approximation has the same second moment as the true Airy pattern.

The quad photodetector associated with lens $i$ lies on the same focal plane and is centered on the lens optical axis. 
We model the quad active area as a square of side length $d_{pq}$ partitioned into four equal square cells. 
Let the total quad support be 
$\Omega_i=\{(x_f,y_f): |x_f|\le d_{pq}/2,\; |y_f|\le d_{pq}/2\}$. 
The four cell regions are then
\begin{align}
	\Omega_{i,1}&=\Big\{(x_f,y_f): 0\le x_f\le \tfrac{d_{pq}}{2},\; 0\le y_f\le \tfrac{d_{pq}}{2}\Big\}, \nonumber\\
	\Omega_{i,2}&=\Big\{(x_f,y_f): -\tfrac{d_{pq}}{2}\le x_f\le 0,\; 0\le y_f\le \tfrac{d_{pq}}{2}\Big\}, \nonumber\\
	\Omega_{i,3}&=\Big\{(x_f,y_f): -\tfrac{d_{pq}}{2}\le x_f\le 0,\; -\tfrac{d_{pq}}{2}\le y_f\le 0\Big\}, \nonumber\\
	\Omega_{i,4}&=\Big\{(x_f,y_f): 0\le x_f\le \tfrac{d_{pq}}{2},\; -\tfrac{d_{pq}}{2}\le y_f\le 0\Big\},
	\label{eq:quad_regions}
\end{align}
as illustrated in Fig.~\ref{cz}(a). 

To capture the post-lens AoA effect per cell, we define the AoA-induced gain of
PD $j\in\{1,2,3,4\}$ behind lens $i$ as the fraction of the Gaussian spot
energy falling into the $j$-th quad region:
\begin{align}
	&h_{\mathrm{AoA},i,j}(\theta_{\mathrm{AoA},x},\theta_{\mathrm{AoA},y}) 
	= \iint_{\Omega_{i,j}}
	\frac{1}{\pi w_{\text{spot}}^2}\nonumber \\
	&~~~\times
	\exp\!\left(
	-\frac{\big\|
		\begin{bmatrix}x_f \\ y_f\end{bmatrix}
		- f_c\begin{bmatrix}\theta_{\mathrm{AoA},x} \\ \theta_{\mathrm{AoA},y}\end{bmatrix}
		\big\|^2}{w_{\text{spot}}^2}
	\right)\,
	dx_f\,dy_f,
	\label{eq:hAoA_int}
\end{align}
where $w_{\text{spot}}$ is the effective spot width, $\mathbf{s}_i=f_c[\theta_{\mathrm{AoA},x},\theta_{\mathrm{AoA},y}]^\top$
is the spot displacement on the focal plane from \eqref{eq:spot_shift}, and
$\Omega_{i,j}$ are the four quad subregions defined in \eqref{eq:quad_regions}.

\paragraph*{Closed-form via $Q(\cdot)$}
The Gaussian PSF in \eqref{eq:gaussian_psf} is separable along $x_f$ and $y_f$, 
which allows closed-form expressions of the quad-cell gains. 
For a general rectangle $[x_L,x_U]\times[y_L,y_U]$, we have
\begin{align}
	&\mathbb{H}_\text{AoA}(x_L,x_U;y_L,y_U) 
	= \Bigg[\Phi\!\left(\frac{x_U-s_{i,x}}{\sigma}\right)
	-\Phi\!\left(\frac{x_L-s_{i,x}}{\sigma}\right)\Bigg]
	\nonumber\\
	&\quad\times
	\Bigg[\Phi\!\left(\frac{y_U-s_{i,y}}{\sigma}\right)
	-\Phi\!\left(\frac{y_L-s_{i,y}}{\sigma}\right)\Bigg],
	\label{eq:rect_int}
\end{align}
where $\Phi(\cdot)$ denotes the standard normal CDF, 
$s_{i,x}=f_c\theta_{\mathrm{AoA},x}$, 
$s_{i,y}=f_c\theta_{\mathrm{AoA},y}$, 
and $\sigma=w_{\text{spot}}/\sqrt{2}$.
Applying \eqref{eq:rect_int} to the four quad regions in \eqref{eq:quad_regions}, 
the per-cell AoA gains are given by
\begin{align}
	h_{\mathrm{AoA},i,1} 
	&= \mathbb{H}_\text{AoA}\!\left(0,\tfrac{d_{pq}}{2};\,0,\tfrac{d_{pq}}{2}\right), \nonumber\\
	h_{\mathrm{AoA},i,2} 
	&= \mathbb{H}_\text{AoA}\!\left(-\tfrac{d_{pq}}{2},0;\,0,\tfrac{d_{pq}}{2}\right), \nonumber\\
	h_{\mathrm{AoA},i,3} 
	&= \mathbb{H}_\text{AoA}\!\left(-\tfrac{d_{pq}}{2},0;\,-\tfrac{d_{pq}}{2},0\right), \nonumber\\
	h_{\mathrm{AoA},i,4} 
	&= \mathbb{H}_\text{AoA}\!\left(0,\tfrac{d_{pq}}{2};\,-\tfrac{d_{pq}}{2},0\right).
	\label{eq:hAoA_closed}
\end{align}

\subsection{Photodetector Output Modeling}
The photocurrent at PD $j$ behind lens $i$, denoted by $y_{i,j}$, can be expressed as
\begin{align}
	y_{i,j} 
	&= P_t \, h_0 \,
	h_{p,i}(\theta_{ex},\theta_{ey}) \,
	h_{\mathrm{AoA},i,j}(\theta_{\mathrm{AoA},x},\theta_{\mathrm{AoA},y}) \,
	h_{a,i}
	+ n_{i,j},
	\label{eq:pd_output}
\end{align}
where $n_{i,j} \sim \mathcal{N}\!\big(0,\sigma_n^2\big)$ is the additive noise, 
$h_0$ denotes the deterministic lumped loss (path and optics),
$h_{a,i}$ is modeled as a Gamma--Gamma distributed random variable with PDF
\begin{align}
	f_{h_{a}}(h)
	&= \frac{2\,(\alpha\beta)^{\frac{\alpha+\beta}{2}}}{\Gamma(\alpha)\,\Gamma(\beta)}
	\, h^{\frac{\alpha+\beta}{2}-1}\,
	K_{\alpha-\beta}\!\left(2\sqrt{\alpha\beta\,h}\right), \quad h>0,
	\label{eq:gg_pdf}
\end{align}
and variance
\begin{align}
	\mathrm{Var}[h_{a,i}] = \frac{1}{\alpha}+\frac{1}{\beta}+\frac{1}{\alpha\beta},
	\label{eq:gg_var}
\end{align}
where $\Gamma(\cdot)$ is the Gamma function and $K_{\nu}(\cdot)$ is the modified Bessel function of the second kind.

\section{Estimation Problem and Learning-Based Inference}

We aim to estimate the random angles and per-aperture atmospheric fades from the quad-PD measurements.
Specifically, the unknown parameter set is
\begin{align}
	\boldsymbol{\Theta}
	&= \big\{\theta_{ex},\,\theta_{ey},\,\theta_{\mathrm{AoA},x},\,\theta_{\mathrm{AoA},y},\,\{h_{a,i}\}_{i=1}^{N_{\text{lens}}}\big\}.
	\label{eq:param_set}
\end{align}
Because the channel is slow-fading, each lens--PD output is collected over a block of
$L_b$ consecutive samples, $\{y_{i,j}[\ell]\}_{\ell=1}^{L_b}$, within which the parameters
in \eqref{eq:param_set} are assumed constant. The block-averaged measurement is
\begin{align}
	\bar{y}_{i,j}
	&= \frac{1}{L_b}\sum_{\ell=1}^{L_b} y_{i,j}[\ell],
	\qquad i=1,\dots,N_{\text{lens}},\; j=1,\dots,4,
	\label{eq:block_avg}
\end{align}
which reduces the additive noise variance as
\begin{align}
	\bar{n}_{i,j} \triangleq \bar{y}_{i,j} - \mathbb{E}[\bar{y}_{i,j}]
	\sim \mathcal{N}\!\left(0,\frac{\sigma_n^2}{L_b}\right).
	\label{eq:noise_reduction}
\end{align}
In practice, $L_b$ cannot be chosen too large, since angular errors must be tracked and corrected rapidly.

Stacking the four PDs of lens $i$ gives
\begin{align}
	\bar{\mathbf{y}}_{i} &= \big[\bar{y}_{i,1},\,\bar{y}_{i,2},\,\bar{y}_{i,3},\,\bar{y}_{i,4}\big]^{\top},
	\label{eq:lens_vec}
\end{align}
and the full measurement vector is
\begin{align}
	\bar{\mathbf{y}} &= \big[\bar{\mathbf{y}}_{1}^{\top},\,\ldots,\,\bar{\mathbf{y}}_{N_{\text{lens}}}^{\top}\big]^{\top}.
	\label{eq:sys_vec}
\end{align}

\subsection{Likelihood and MAP Estimation}
Using the PD model in \eqref{eq:pd_output} and the block-averaging in \eqref{eq:block_avg}--\eqref{eq:noise_reduction},
the Gaussian likelihood of $\bar{\mathbf{y}}$ given $\boldsymbol{\Theta}$ is
\begin{align}
	p(\bar{\mathbf{y}}\,|\,\boldsymbol{\Theta})
	&\propto \exp\!\Bigg(
	-\frac{L_b}{2\sigma_n^2}\!
	\sum_{i=1}^{N_{\text{lens}}}\sum_{j=1}^{4}
	\big|\bar{y}_{i,j} - \mu_{i,j}(\boldsymbol{\Theta})\big|^2
	\Bigg),
	\label{eq:likelihood}
\end{align}
where
\begin{align}
	\mu_{i,j}(\boldsymbol{\Theta})
	&= P_t \, h_0 \,
	h_{p,i}(\theta_{ex},\theta_{ey}) \,
	h_{\mathrm{AoA},i,j}(\theta_{\mathrm{AoA},x},\theta_{\mathrm{AoA},y}) \,
	h_{a,i}.
	\label{eq:mu_ij}
\end{align}
With the Gamma--Gamma prior $p(h_{a,i})$ of \eqref{eq:gg_pdf} (independent across $i$),
a MAP estimate solves
\begin{align}
	\hat{\boldsymbol{\Theta}}_{\mathrm{MAP}}
	&= \arg\min_{\boldsymbol{\Theta}}
	\Bigg\{
	\frac{L_b}{2\sigma_n^2}
	\sum_{i=1}^{N_{\text{lens}}}\sum_{j=1}^{4}
	\big|\bar{y}_{i,j} - \mu_{i,j}(\boldsymbol{\Theta})\big|^2
	\nonumber\\[-2pt]
	&\qquad\qquad\qquad\quad
	- \sum_{i=1}^{N_{\text{lens}}}\log p\big(h_{a,i}\big)
	\Bigg\}.
	\label{eq:map_obj}
\end{align}
The objective in \eqref{eq:map_obj} couples the four angles through
$h_{p,i}(\cdot)$ and $h_{\mathrm{AoA},i,j}(\cdot)$, while promoting physically
plausible $\{h_{a,i}\}$ via the Gamma--Gamma prior.

\subsection{Deep Learning-Based Estimation Framework}
Motivated by the nonlinear coupling between the pointing errors, the 
AoA fluctuations, and the turbulence-induced fading, 
we employ a data-driven estimation approach based on DL. 
Unlike classical model-based estimators, which require explicit derivation 
of closed-form likelihoods in \eqref{eq:likelihood} and iterative optimization 
of the MAP objective in \eqref{eq:map_obj}, a DL-based framework can directly 
learn the highly nonlinear mapping from the measured PD outputs 
$\bar{\mathbf{y}}$ in \eqref{eq:sys_vec} to the underlying channel parameters. 

Specifically, we design a fully-connected multilayer perceptron (MLP) to 
approximate the unknown function
\begin{align}
	f_{\boldsymbol{\Theta}}: \bar{\mathbf{y}} \;\mapsto\; 
	\big[\theta_{ex},\,\theta_{ey},\,\theta_{\mathrm{AoA},x},\,\theta_{\mathrm{AoA},y},
	\{h_{a,i}\}_{i=1}^{N_{\text{lens}}}\big],
	\label{eq:mlp_mapping}
\end{align}
where the network learns the joint estimation of both angular errors and 
per-aperture turbulence coefficients. The universal approximation property 
of neural networks guarantees that, given sufficient training samples, the 
MLP can closely approximate the MAP-optimal estimator in \eqref{eq:map_obj}. 

The scientific rationale for adopting a DL model is twofold. 
First, the nonlinearities arising from the multiplicative structure of 
$h_{p,i}(\cdot)$ and $h_{\mathrm{AoA},i,j}(\cdot)$ in \eqref{eq:mu_ij} 
make analytical solutions intractable, especially under Gamma--Gamma fading. 
Second, real-world impairments such as hardware distortions or unmodeled 
jitter cannot be fully captured by analytical channel models, whereas a 
data-driven DL estimator can inherently learn such effects from training data. 
Consequently, the DL framework provides both robustness and scalability, 
allowing real-time inference once the network is trained.

\subsection{Proposed Hierarchical Learning Approach}
Although the end-to-end DL estimator in \eqref{eq:mlp_mapping} is universal, 
it faces three main challenges: (i) simultaneous learning of multiple nonlinear 
dependencies increases training complexity and inference latency; 
(ii) the model may overfit to specific channel conditions and fail to 
generalize under mobile FSO topologies; and (iii) it requires a very large 
dataset to cover the full parameter space. 

To address these issues, we propose a \emph{hierarchical learning} strategy 
that decomposes the task into sequential stages. Since different physical 
impairments leave distinct imprints on the received signal 
$\bar{\mathbf{y}}$ in \eqref{eq:sys_vec}, estimating them step by step 
reduces the learning burden, improves interpretability, and enhances robustness.

\subsubsection{Stage~1: AoA Estimation}

At the first stage, our goal is to estimate the instantaneous 
AoA vector 
$\boldsymbol{\theta}_{\mathrm{AoA}} = 
(\theta_{\mathrm{AoA},x},\theta_{\mathrm{AoA},y})$ defined in 
\eqref{eq:aoa_def}. The AoA information is primarily encoded in the 
\emph{relative distribution of energy} across the four quadrants of each 
quad photodetector, whereas the absolute power is affected by pointing 
loss $h_{p,i}(\cdot)$ and turbulence fading $h_{a,i}$. To reduce the 
impact of these multiplicative factors, we normalize the four outputs 
of lens $i$ as
\begin{align}
	\tilde{y}_{i,j}
	&= \frac{\bar{y}_{i,j}}
	{\sum_{k=1}^{4}\bar{y}_{i,k}},
	\qquad j=1,\ldots,4,
	\label{eq:quad_norm}
\end{align}
By construction, the normalized vector 
$\tilde{\mathbf{y}}_{i}=[\tilde{y}_{i,1},\ldots,\tilde{y}_{i,4}]^\top$ 
satisfies $\sum_{j=1}^4 \tilde{y}_{i,j}\approx 1$, and is largely 
independent of the common scaling due to $h_{p,i}$ and $h_{a,i}$. 
Consequently, the residual variations in $\tilde{\mathbf{y}}_{i}$ 
are dominated by the AoA-induced spot displacement in 
\eqref{eq:hAoA_closed}. 

A lightweight MLP is then trained to map 
$\{\tilde{\mathbf{y}}_{i}\}_{i=1}^{N_{\text{lens}}}$ into an AoA estimate
\begin{align}
	\hat{\boldsymbol{\theta}}_{\mathrm{AoA}}
	= f_{\mathrm{MLP}}^{(1)}\big(\{\tilde{\mathbf{y}}_{i}\}\big),
	\label{eq:aoa_mlp}
\end{align}
which provides a robust estimate of the angular jitter despite 
turbulence and pointing fluctuations. 

After obtaining $\hat{\boldsymbol{\theta}}_{\mathrm{AoA}}$, the four PD 
outputs of each lens are aggregated to form the total received signal per lens
\begin{align}
	\bar{y}_{i}^{\Sigma} 
	&= \sum_{j=1}^{4} \bar{y}_{i,j},
	\label{eq:y_sum}
\end{align}
while the corresponding total AoA gain is computed as
\begin{align}
	\hat{h}^{\mathrm{tot}}_{\mathrm{AoA},i} 
	&= \sum_{j=1}^{4} 
	h_{\mathrm{AoA},i,j}\!\left(\hat{\theta}_{\mathrm{AoA},x},\hat{\theta}_{\mathrm{AoA},y}\right).
	\label{eq:hAoA_tot}
\end{align}
The normalized per-lens signal for the next stage is then defined by 
removing the AoA contribution multiplicatively:
\begin{align}
	z_i^{(1)}
	&= \frac{\bar{y}_{i}^{\Sigma}}
	{P_t h_0 \,\hat{h}^{\mathrm{tot}}_{\mathrm{AoA},i} }
	\;\approx\;
	h_{p,i}(\theta_{ex},\theta_{ey})\,h_{a,i}.
	\label{eq:stage1_div}
\end{align} 
Thus, Stage~1 not only yields the AoA estimate in 
\eqref{eq:aoa_mlp}, but also produces the per-lens signal $z_i^{(1)}$ 
which is largely free of AoA distortions and ready for Stage~2 estimation.

\subsubsection{Stage~2: Transmitter Pointing Error Estimation}
From \eqref{eq:stage1_div} and \eqref{eq:hpi_full}, each entry follows
the multiplicative structure
\begin{align}
	z^{(1)}_i \;=\; h_{p,i}(\boldsymbol{\theta}_e)\, h_{a,i} \;+\; \nu_i,
	\qquad i=1,\ldots,N_{\text{lens}},
	\label{eq:z1_struct}
\end{align}
where $h_{p,i}(\cdot)$ encodes the geometric pointing pattern across the 
lens array and $\{h_{a,i}\}$ are i.i.d.\ Gamma--Gamma fades (cf. \eqref{eq:gg_pdf}). 
Crucially, the \emph{spatial} envelope of $\mathbf{z}^{(1)}$ is governed by 
$h_{p,i}(\boldsymbol{\theta}_e)$, while $\{h_{a,i}\}$ act as independent 
per-lens scalings.

To extract $\boldsymbol{\theta}_e=(\theta_{ex},\theta_{ey})$, we employ a 
dedicated neural estimator that directly maps the vector of per-lens signals 
to the angular error:
\begin{align}
	\hat{\boldsymbol{\theta}}_e
	\;=\; f_{\mathrm{MLP}}^{(2)}\!\big(\mathbf{z}^{(1)}\big),
	\label{eq:theta_e_mlp}
\end{align}
where $f_{\mathrm{MLP}}^{(2)}(\cdot)$ is a fully-connected network trained 
to approximate the nonlinear maximum-likelihood mapping implied by 
\eqref{eq:z1_struct}. The input dimension scales with $N_{\text{lens}}$, so 
a denser lens array provides more spatial observations of the pointing 
profile, enabling the MLP to achieve higher estimation accuracy.

\subsubsection{Stage~3: Receiver Jitter and Turbulence Estimation}
Given the Stage-1 AoA estimate $\hat{\boldsymbol{\theta}}_{\mathrm{AoA}}$ in \eqref{eq:aoa_mlp}
and the Stage-2 transmitter pointing estimate $\hat{\boldsymbol{\theta}}_e$ in \eqref{eq:theta_e_mlp},
the receiver-induced angular jitter follows directly from the additive AoA model in \eqref{eq:aoa_def}:
\begin{align}
	\hat{\boldsymbol{\theta}}_r
	\;=\;
	\hat{\boldsymbol{\theta}}_{\mathrm{AoA}} \;-\; \hat{\boldsymbol{\theta}}_e.
	\label{eq:theta_r_hat}
\end{align}
Equation \eqref{eq:theta_r_hat} separates the platform jitter from the transmitter tracking error
using purely learned angular estimates, without requiring any additional sensing.

Starting from \eqref{eq:z1_struct}, once $\hat{\boldsymbol{\theta}}_e$ is available, the
per-lens turbulence coefficients are obtained by multiplicative pointing removal:
\begin{align}
	\hat{h}_{a,i}
	\;=\;
	\frac{z_i^{(1)}}{\,h_{p,i}\!\big(\hat{\boldsymbol{\theta}}_e\big)\,},
	\qquad i=1,\ldots,N_{\text{lens}}.
	\label{eq:ha_simple}
\end{align}

\begin{table*}[t]
	\centering
	\caption{Estimated Angles (µrad) and Turbulence Coefficients for Different Methods}
	\begin{tabular}{|c|c|rrrrrrrr|}
		\hline
		Sample & Method & $\theta_{ex}$ & $\theta_{ey}$ & $\theta_{rx}$ & $\theta_{ry}$ & $\theta_{\text{AoA},x}$ & $\theta_{\text{AoA},y}$ & $h_{a,1}$ & $h_{a,2}$ \\
		\hline
		& True          &  -55 & -496 & 3396 & -1075 & 3340 & -1571 & 1.844 & 0.602 \\
		1 & MAP           &  -58 & -510 & 3390 & -1006 & 3332 & -1516 & 1.840 & 0.603 \\
		& End2End       &  -63 & -505 & 3306 & -1251 & 3243 & -1756 & 1.842 & 0.603 \\
		& Hierarchical  &  -49 & -482 & 3315 & -1042 & 3266 & -1524 & 1.849 & 0.602 \\
		\hline\hline
		& True          &   84 &  -53 & 2832 &  -705 & 2916 &  -759 & 0.691 & 0.454 \\
		2 & MAP           &   83 &  -58 & 2764 &  -712 & 2847 &  -770 & 0.690 & 0.453 \\
		& End2End       &   81 &  -57 & 3034 &  -707 & 3114 &  -764 & 0.690 & 0.453 \\
		& Hierarchical  &   80 &  -52 & 2741 &  -721 & 2821 &  -773 & 0.691 & 0.453 \\
		\hline\hline
		& True          &  140 &  833 & -2775 & 1652 & -2635 & 2485 & 0.992 & 1.907 \\
		3 & MAP           &  135 &  793 & -2730 & 1724 & -2595 & 2517 & 0.963 & 1.866 \\
		& End2End       &  143 &  815 & -2843 & 1622 & -2700 & 2438 & 0.981 & 1.892 \\
		& Hierarchical  &  144 &  937 & -2799 & 1366 & -2655 & 2303 & 1.074 & 2.021 \\
		\hline
	\end{tabular}
	\label{tab:estimation_results}
\end{table*}

\begin{table*}[t]
	\centering
	\caption{Normalized MSE (NMSE) of angle and turbulence estimates for different lens array sizes}
	\begin{tabular}{|c|c|c c c c c c c c|}
		\hline
		\multicolumn{2}{|c|}{} & \multicolumn{8}{c|}{NMSE} \\
		\hline
		$N_{\text{lens}}$ & Method & $\theta_{ex}$ & $\theta_{ey}$ & $\theta_{rx}$ & $\theta_{ry}$ & $\theta_{\mathrm{AoA},x}$ & $\theta_{\mathrm{AoA},y}$ & $h_{a,1}$ & $h_{a,2}$ \\
		\hline
		4  & End2End       & 0.2330 & 0.2299 & 0.0833 & 0.0828 & 0.0092 & 0.0092 & $24.2$ & $25.3$ \\
		& Hierarchical  & 0.2869 & 0.2918 & 0.0978 & 0.0999 & 0.0037 & 0.0037 & $3.08{\times}10^{7}$ & $4.51{\times}10^{8}$ \\
		\hline
		16 & End2End       & 0.0144 & 0.0143 & 0.0065 & 0.0065 & 0.0025 & 0.0026 & 0.0386 & 0.0406 \\
		& Hierarchical  & 0.0188 & 0.0183 & 0.0071 & 0.0068 & 0.0010 & 0.0010 & 0.0650 & 0.0580 \\
		\hline
		64 & End2End       & 0.0016 & 0.0016 & 0.0011 & 0.0011 & 0.0009 & 0.0009 & 0.0030 & 0.0030 \\
		& Hierarchical  & 0.0020 & 0.0020 & 0.0009 & 0.0009 & 0.0004 & 0.0004 & 0.0037 & 0.0035 \\
		\hline
	\end{tabular}
	\label{tab:nmse_results_lens}
\end{table*}

\section{Simulation Parameters}
Unless stated otherwise, the default parameters are as follows. 
We generate $2\times 10^6$ i.i.d.\ channel realizations per setup, 
half used for training and half for testing. 
The optical wavelength is $\lambda=1550$~nm, link distance $Z_L=500$~m, 
divergence $\theta_{\mathrm{div}}=2$~mrad, beam waist 
$w_z=\theta_{\mathrm{div}} Z_L=1$~m, and effective focal-spot width 
$w_{\text{spot}}=400~\mu$m. 
Each small lens has radius $r_a=2$~cm and focal length $f_c=4$~cm. 
The array size is varied as $N_{\text{lens}}\!\in\!\{4,16,64\}$, 
with inter-lens pitch $5r_a$. 
The quad-PD has total side length $d_{pq}=4$~mm. 
Transmit power is $P_t=0.1$~W with deterministic loss $h_0=0.7$. 
The transmitter pointing error is Gaussian with 
$\sigma_\theta=1$~mrad per axis; receiver jitter 
has $\sigma_r=1$~mrad. 
Atmospheric fading follows the Gamma--Gamma distribution in \eqref{eq:gg_pdf} 
with $(\alpha,\beta)=(8,6)$. 
Additive Gaussian noise has variance $\sigma_n^2=(10^{-7})^2$ per PD sample; 
block averaging with $L_b=1000$. 

For deep learning, we use fully-connected MLPs. 
In single-task setups (e.g., only $\boldsymbol{\theta}_e$ or AoA), 
compact networks with two hidden layers ($128\!-\!64$ neurons, Leaky-ReLU) suffice. 
For the end-to-end baseline \eqref{eq:mlp_mapping}, we adopt a larger architecture 
($512\!-\!256\!-\!128$ hidden units) to handle the joint mapping from 
$\bar{\mathbf{y}}$ to all parameters. 
Training on $10^6$ samples requires $\sim$1~h for the end-to-end model 
and $\sim$0.5~h for the hierarchical pipeline (all stages), 
on a single modern GPU. For comparison, MAP estimation \eqref{eq:map_obj} is evaluated 
by an exhaustive search over $10^8$ candidate states, 
which provides a near-oracle reference but is computationally prohibitive 
for practical deployment.

The results in Table~\ref{tab:estimation_results} correspond to the case $N_{\text{lens}}=16$. 
It can be observed that the MAP estimator provides the most accurate estimates across all parameters. 
However, MAP requires an exhaustive search over a very large candidate space, which makes it computationally 
infeasible for real-time FSO systems. In contrast, both deep learning methods can be practically deployed 
with manageable complexity. Between the two, the hierarchical framework achieves noticeably better accuracy 
for AoA estimation, while the end-to-end model tends to perform better for transmitter pointing error estimation.

For a more comprehensive performance assessment, it is more appropriate to evaluate the normalized MSE 
(NMSE) over a large number of Monte-Carlo realizations. Since MAP evaluation at such scale is not tractable, 
in Table~\ref{tab:nmse_results_lens} we compare only the end-to-end and the hierarchical approaches 
for different lens array sizes. The results reveal that AoA can be accurately estimated even with a small number 
of lenses, and in this regime the hierarchical model provides better robustness. In contrast, estimation of the 
transmitter pointing error $\boldsymbol{\theta}_e$ is more demanding, as the pointing gain $h_{p,i}(\cdot)$ 
in~\eqref{eq:hpi_full} is multiplicatively coupled with the turbulence coefficient $h_{a,i}$. This coupling 
acts as an additional source of distortion, and with a small number of lenses the diversity of spatial 
observations is insufficient, leading to degraded performance in both methods. By increasing the lens array size, 
the spatial diversity compensates for this loss, and both models can reliably estimate $\boldsymbol{\theta}_e$. 
In this regime, the end-to-end model achieves slightly higher accuracy, but requires significantly larger 
training data and more complex neural architectures. The hierarchical pipeline, on the other hand, 
learns faster and provides competitive accuracy with substantially lower training complexity.

Finally, the turbulence coefficients $\{h_{a,i}\}$ are strongly tied to the transmitter pointing accuracy, 
since any small estimation error in $\boldsymbol{\theta}_e$ directly propagates into the pointing gain and hence 
into the turbulence estimate. Therefore, precise transmitter angle estimation is critical to avoid large 
errors in turbulence reconstruction. The results in Table~\ref{tab:nmse_results_lens} confirm that, once 
$\boldsymbol{\theta}_e$ is reliably estimated with a sufficient number of lenses, both learning strategies 
can achieve accurate turbulence coefficient estimation.

\bibliographystyle{IEEEtran}
\bibliography{IEEEabrv,myref}

\begin{thebibliography}{10}
\providecommand{\url}[1]{#1}
\csname url@samestyle\endcsname
\providecommand{\newblock}{\relax}
\providecommand{\bibinfo}[2]{#2}
\providecommand{\BIBentrySTDinterwordspacing}{\spaceskip=0pt\relax}
\providecommand{\BIBentryALTinterwordstretchfactor}{4}
\providecommand{\BIBentryALTinterwordspacing}{\spaceskip=\fontdimen2\font plus
\BIBentryALTinterwordstretchfactor\fontdimen3\font minus
  \fontdimen4\font\relax}
\providecommand{\BIBforeignlanguage}[2]{{%
\expandafter\ifx\csname l@#1\endcsname\relax
\typeout{** WARNING: IEEEtran.bst: No hyphenation pattern has been}%
\typeout{** loaded for the language `#1'. Using the pattern for}%
\typeout{** the default language instead.}%
\else
\language=\csname l@#1\endcsname
\fi
#2}}
\providecommand{\BIBdecl}{\relax}
\BIBdecl

\bibitem{kumar2025skr}
S.~Kumar and S.~P. Dash, ``{SKR Analysis of One- and Two-Way CV-QKD MIMO FSO
  Communication System},'' \emph{IEEE Communications Letters}, 2025.

\bibitem{xu2025multi}
F.~Xu \emph{et~al.}, ``{Multi-Antenna UAV Assisted Hybrid FSO/RF Data
  Collection for IoT: Optimal Design for Fairness},'' \emph{IEEE Transactions
  on Aerospace and Electronic Systems}, 2025.

\bibitem{10681506}
M.~T. Dabiri, M.~Hasna, S.~Althunibat, and K.~Qaraqe, ``{Modulating
  Retroreflector-Based Satellite-to-Ground Optical Links: Joint Communications
  and Tracking},'' \emph{IEEE Transactions on Communications}, vol.~73, no.~3,
  pp. 1950--1962, 2025.

\bibitem{10553228}
M.~Taghi~Dabiri, M.~Hasna, S.~Althunibat, and K.~Qaraqe, ``{Modulating
  Retroreflector-Based Satellite-to-Ground Optical Communications: Acquisition,
  Sensing, and Positioning},'' \emph{IEEE Transactions on Communications},
  vol.~73, no.~1, pp. 483--497, 2025.

\bibitem{paul2022pulse}
P.~Paul, M.~R. Bhatnagar, and J.~Nebhen, ``{Pulse Jamming in Aperture-Averaged
  FSO Receiver Over Exponentiated Weibull Fading Channel},'' \emph{IEEE
  Transactions on Wireless Communications}, vol.~21, no.~6, pp. 4242--4254,
  June 2022.

\bibitem{girdher2024ris}
A.~Girdher and A.~Bansal, ``{RIS-Assisted Multi-Aperture FSO Communication
  Network for High-Speed Train: Second-Order Statistical Analysis},''
  \emph{IEEE Transactions on Intelligent Transportation Systems}, vol.~25,
  no.~10, pp. 14\,140--14\,153, Oct. 2024.

\bibitem{ata2023haps}
Y.~Ata and M.-S. Alouini, ``{HAPS Based FSO Links Performance Analysis and
  Improvement With Adaptive Optics Correction},'' \emph{IEEE Transactions on
  Wireless Communications}, vol.~22, no.~7, pp. 4916--4929, July 2023.

\bibitem{solanki2025computationally}
P.~B. Solanki, S.~D. Bopardikar, and X.~Tan, ``{Computationally Efficient
  Control for Cooperative Optical Beam Tracking With Guaranteed Finite-Time
  Convergence},'' \emph{IEEE Transactions on Control Systems Technology},
  vol.~33, no.~1, pp. 245--260, Jan. 2025.

\bibitem{chen2021parameter}
D.~Chen and J.~Hui, ``{Parameter estimation of Gamma--Gamma fading channel in
  free space optical communication},'' \emph{Optics Communications}, vol. 488,
  p. 126830, June 2021.

\bibitem{abdavinejad2020fso}
H.~Abdavinejad, E.~Mostafapour, R.~F. Garaboghloo, C.~Ghobadi, J.~Nourinia, and
  A.~Soleimani, ``{Free space optical (FSO) channel estimation and tracking
  using the incremental adaptive networks},'' in \emph{2020 3rd West Asian
  Symposium on Optical and Millimeter-wave Wireless Communication
  (WASOWC)}.\hskip 1em plus 0.5em minus 0.4em\relax IEEE, Nov. 2020, pp. 1--5.

\bibitem{kim2025moment}
Y.~Kim and D.~Yoon, ``{Moment-Based Estimation for Gamma-Gamma Fading
  Parameters in Free-Space Optical Links},'' \emph{IEEE Journal on Selected
  Areas in Communications}, vol.~43, no.~5, pp. 1582--1589, May 2025.

\bibitem{han2022joint}
L.~Han, X.~Liu, Y.~Wang, and B.~Li, ``{Joint Impact of Channel Estimation
  Errors and Pointing Errors on FSO Communication Systems Over $\mathcal{F}$
  Turbulence Channel},'' \emph{Journal of Lightwave Technology}, vol.~40,
  no.~14, pp. 4555--4561, July 2022.

\bibitem{dhanalakshmi2022onboard}
R.~Dhanalakshmi \emph{et~al.}, ``{Onboard Pointing Error Detection and
  Estimation of Observation Satellite Data Using Extended Kalman Filter},''
  \emph{Computational Intelligence and Neuroscience}, vol. 2022, pp. 1--8, Oct.
  2022.

\bibitem{miao2021parameter}
M.~Miao, W.~Cai, and X.~Li, ``{Parameter Estimation of Gamma-Gamma Fading with
  Generalized Pointing Errors in FSO Systems},'' \emph{Wireless Communications
  and Mobile Computing}, vol. 2021, no.~1, Jan. 2021.

\bibitem{dreier2025beam}
H.~F. Dreier, A.~Derakhshandeh, A.~Harlakin, and P.~A. Hoeher,
  ``{Beam-Shape-Based Angle-of-Arrival Estimation in Free-Space Laser
  Communication},'' \emph{IEEE Photonics Journal}, vol.~17, no.~2, pp. 1--16,
  Apr. 2025.

\bibitem{tsai2023angle}
M.-C. Tsai, M.~S. Bashir, and M.-S. Alouini, ``{Angle-of-Arrival Estimation of
  Narrow Gaussian Beams for Mobile FSO Platforms},'' \emph{arXiv preprint
  arXiv:2307.16002}, 2023.

\bibitem{dabiri2025novel}
M.~T. Dabiri and M.~Hasna, ``{A Novel MRR-UAV based Relay with Optical Network
  Coding: A Comparative Study with Optical IRS and Conventional UAV
  Relaying},'' \emph{IEEE Journal on Selected Areas in Communications}, 2025.

\end{thebibliography}

\end{document}